# Wave Mechanics without Probability

Adriano Orefice*, Raffaele Giovanelli, Domenico Ditto

*Università degli Studi di Milano - DISAA - Via Celoria, 2 - 20133 - Milano (Italy)*

**ABSTRACT.** The behavior of monochromatic electromagnetic waves in stationary media is shown to be ruled by a *frequency dependent* function, which we call *Wave Potential*, encoded in the structure of the Helmholtz equation. *Contrary to the common belief that the very concept of "ray trajectory" is reserved to the eikonal approximation,* a general and exact *ray-based* Hamiltonian treatment, reducing to the eikonal approximation in the *absence* of Wave Potential, shows that its *presence* induces a mutual, perpendicular *ray-coupling*, which is *the one and only cause* of any typically wave-like phenomenon, such as *diffraction and interference*.

Recalling, then, that the **time-independent** Schrödinger and Klein-Gordon equations (associating *stationary "matter waves"* to mono-energetic particles) are themselves Helmholtz-like equations, the exact, ray-based treatment developed for classical electromagnetic waves is extended - *without resorting to statistical concepts* - to the exact, *trajectory-based* Hamiltonian dynamics of mono-energetic *point-like particles*, both in the non-relativistic and in the relativistic case. The trajectories turn out to be perpendicularly coupled, once more, by an *exact, stationary, energy-dependent* Wave Potential, coinciding in the form, but not in the physical meaning, with the *statistical, time-varying, energy-independent* "Quantum Potential" of Bohm's theory, which views particles, just like the standard Copenhagen interpretation, as traveling *wave-packets*. These results, together with the connection which is shown to exist between Wave Potential and Uncertainty Principle, suggest a novel, *non-probabilistic* interpretation of Wave Mechanics.

**KEYWORDS.** *Wave Mechanics - de Broglie's theory - Matter waves - Pilot waves - Wave equation - Helmholtz equation - Hamilton equations - Hamilton-Jacobi equations - Schrödinger equation - Klein-Gordon equation - Wave potential - Quantum potential - Classical dynamics - Quantum dynamics - Bohm's theory - Electromagnetic waves - Ray trajectories - Quantum trajectories - Uncertainty Principle - Eikonal approximation - Geometrical optics approximation - Wave diffraction - Wave interference*

PACS 03.50.De  *Classical Electromagnetism. Maxwell equations*
PACS 03.75.-b  *Matter Waves*
PACS 03.65.-w  *Quantum mechanics*
PACS 03.65.Ta  *Foundations of quantum mechanics. Measurement theory*
PACS 03.30.+p  *Special Relativity*

## 1 – Introduction

*Let me say at the outset that I am opposing not a few special statements of quantum physics held today (1955): I am opposing the whole of it (...), I am opposing its basic views, shaped when Max Born put forward his **probabilistic interpretation**, which was accepted by almost everybody.* (E. Schrödinger [**1**])

Let us say at the outset that we put forward in the present paper, borrowing its title from Everett's famous "*long thesis*"[**2**], **a non-probabilistic interpretation** of Wave Mechanics adverse both to Born's and Everett's standpoints. Our interpretation is based on a line of research [**3-6**] starting from the demonstration, for the first time in the development of *Classical Mechanics*, that any kind of wave-like features may be treated (for monochromatic waves described by a Helmholtz-like equation) by means of a *ray-based treatment*: a property which was previously thought to be reserved to the geometrical optics approximation.

---

* Corresponding author - adriano.orefice@unimi.it

This treatment (expressed in terms of an *exact* Hamiltonian kinematics) is allowed by the discovery of a *dispersive function* which we call *Wave Potential*, and which we show:

- to be encoded in the structure of the Helmholtz equation,
- to cause a mutual perpendicular coupling between monochromatic rays,
- to avoid any statistical concept, and
- to be the one and only cause of any typically wave-like feature, such as diffraction and interference, while its omission leads to the usual geometrical optics approximation.

Analogous considerations are then extended to the case of *Wave Mechanics*, thanks to the fact that both the *time-independent* Schrödinger and Klein-Gordon equations (associating *mono-energetic* material particles with suitable, stationary "*matter waves*") are themselves Helmholtz-like equations, allowing to formulate a self-consistent *Hamiltonian dynamics of point-like particles* in terms of *exact trajectories and motion laws*, under the *piloting* rule of a *Wave Potential* function, in whose absence they reduce to the usual laws of *classical dynamics*.

Let us remind here for comparison that, according to Bohm's theory [**7**, **8**], "*the use of statistical ensembles, although not a reflection of an inherent limitation on the precision with which it is correct for us to conceive of the variables defining the state of the system, is a **practical necessity**, as in the case of classical statistical mechanics*". Bohm's approach represents therefore particles by means of statistical **wave-packets** (traveling according to the **time-dependent Schrödinger equation**) just like in the orthodox Copenhagen interpretation, to which it associates a set of probability flow-lines [**9-12**]: a result which gave rise, by itself, to extensive applications, ranging from chemical physics to nanoscale systems [**13**-**17**].

In our approach, on the other hand, Bohm's **practical necessity** is bypassed by a set of *exact* (i.e. *non-statistical*) dynamical equations, directly stemming from the **time-independent Schrödinger equation** and providing the exact, trajectory-based dynamics of mono-energetic **point-like particles**, without requiring the simultaneous solution of *time-dependent* Schrödinger (or Klein-Gordon) equations. The particles follow, in other words - *just like in classical dynamics* - a system of exact trajectories, of which Bohm's flow-lines represent a *statistical average*.

We sketch in Sects.2 and 3 the definition and role of the Wave Potential, presenting in Sect.4 a few considerations about the connection between our non-probabilistic trajectories and the Uncertainty Principle. We extend in Sect.5 our approach to the relativistic case, and stress in Sect.6 the physical difference between Bohm's *unavoidably statistical* theory and our *exact* one. We finally draw physical conclusions in Sect.7 and 8, suggesting a new, *non-probabilistic* approach to Wave Mechanics.

**2 - The Wave Potential function in Classical Mechanics**

By assuming both *wave monochromaticity* and *stationary, isotropic media* we briefly sketch here our approach [**3-6, 18**] holding, in principle, for any kind of *classical* waves described by Helmholtz-like equations. In order to fix ideas we refer, within the present Section, to the case of *classical* electromagnetic waves of the form

$$\psi(\vec{r},\omega,t) = u(\vec{r},\omega)\ e^{-i\omega t}\ , \qquad (1)$$

where $\vec{r} \equiv (x,y,z)$, and $u(\vec{r},\omega)$ is a solution of the Helmholtz equation [**19**]

$$\nabla^2 u + (n k_0)^2 u = 0, \qquad (2)$$

where $\nabla^2 \equiv \partial^2/\partial x^2 + \partial^2/\partial y^2 + \partial^2/\partial z^2$ ; $k_0 \equiv \dfrac{2\pi}{\lambda_0} = \dfrac{\omega}{c}$ ; the function $\psi(\vec{r},\omega,t)$ represents any component of the electric and/or magnetic field of the wave, and $n(\vec{r},\omega)$ is the (time independent) refractive index of the medium. If we now perform the (quite general) replacement [19]

$$u(\vec{r},\omega) = R(\vec{r},\omega) \, e^{i \, \varphi(\vec{r},\omega)}, \qquad (3)$$

with real $R(\vec{r},\omega)$ and $\varphi(\vec{r},\omega)$, where $R(\vec{r},\omega)$ represents, *without any intrinsically probabilistic meaning*, the space distribution of the monochromatic wave amplitude, we obtain, after separation of real and imaginary parts and the definition of the wave vector

$$\vec{k} = \vec{\nabla} \varphi(\vec{r},\omega), \qquad (4)$$

with $\vec{\nabla} \equiv \partial/\partial \vec{r} \equiv (\partial/\partial x, \partial/\partial y, \partial/\partial z)$, the equation systems

$$\vec{\nabla} \cdot (R^2 \vec{\nabla} \varphi) = 0 \qquad (5)$$

$$(\vec{\nabla} \varphi)^2 - (n k_0)^2 = -\frac{2 k_0}{c} W(\vec{r},\omega) \qquad (6)$$

$$\begin{cases} \dfrac{d\vec{r}}{dt} = \dfrac{\partial D}{\partial \vec{k}} \equiv \dfrac{c\vec{k}}{k_0} & (7) \\ \dfrac{d\vec{k}}{dt} = -\dfrac{\partial D}{\partial \vec{r}} \equiv \vec{\nabla}[\dfrac{c k_0}{2} n^2(\vec{r},\omega) - W(\vec{r},\omega)] & (8) \end{cases}$$

with $\partial/\partial \vec{k} \equiv (\partial/\partial k_x, \partial/\partial k_y, \partial/\partial k_z)$ and

$$W(\vec{r},\omega) = -\frac{c}{2 k_0} \frac{\nabla^2 R(\vec{r},\omega)}{R(\vec{r},\omega)} \quad ; \quad D(\vec{r},\vec{k},\omega) = \frac{c}{2 k_0}[k^2 - (n k_0)^2 + W(\vec{r},\omega)] \quad . \qquad (9)$$

The Hamiltonian system (7)-(8), satisfying the differentiation $\dfrac{\partial D}{\partial \vec{r}} \cdot d\vec{r} + \dfrac{\partial D}{\partial \vec{k}} \cdot d\vec{k} = 0$ allowed by eq.(6), associates an **exact kinematical ray-tracing** to the Helmholtz equation (2). A ray velocity $\vec{v}_{ray} = \dfrac{c\vec{k}}{k_0}$ is implicitly defined, and we may notice that, as long as $k \equiv |\vec{k}|$ remains equal to its launching value $k_0$, we'll have $v_{ray} \equiv |\vec{v}_{ray}| = c$.

The function $W(\vec{r},\omega)$, called here "*Helmholtz Wave Potential*" (although it has the dimensions of a *frequency*), plays the basic role of mutually coupling the ray-trajectories relevant to the considered monochromatic wave, in a kind of self-refraction (which we call "*Helmholtz coupling*") affecting both their geometry and their motion laws. Eq.(5), expressing the constancy of the flux of the vector field $R^2 \vec{\nabla} \varphi$ along any tube formed by the field lines of the wave-vector $\vec{k} = \vec{\nabla} \varphi(\vec{r},\omega)$, may be written in the form $\vec{\nabla} \cdot (R^2 \vec{\nabla} \varphi) \equiv 2 R \vec{\nabla} R \cdot \vec{\nabla}\varphi + R^2 \vec{\nabla} \cdot \vec{\nabla}\varphi = 0$, and plays a double role:

- <u>On the one hand</u>, since no new trajectory may suddenly arise in the space region spanned by the considered wave trajectories, we must have $\vec{\nabla} \cdot \vec{\nabla}\varphi = 0$, so that $\vec{\nabla} R \cdot \vec{\nabla} \varphi = 0$: the amplitude $R(\vec{r},\omega)$, together with its derivatives and functions, including $W(\vec{r},\omega)$, is distributed, at any step of the numerical integration of eqs.(7)-(8), over the wave-front, normal at each point to $\vec{k} \equiv \vec{\nabla} \varphi(\vec{r},\omega)$, reached at that step, so that $\vec{\nabla} W(\vec{r},\omega) \cdot \vec{k} = 0$, and the "Helmholtz coupling" acts *perpendicularly* to the wave trajectories. A fundamental consequence of this property is the fact that, in the case of electromagnetic waves propagating *in vacuo* (i.e. for $n = 1$), the absolute value of the ray velocity $\vec{v}_{ray} = \dfrac{c\,\vec{k}}{k_0}$, as shown by eq.(8), remains equal to $c$ all along each ray trajectory, whatever its form may be;

- <u>on the other hand</u>, the Hamiltonian system (7)-(8) is consistently *"closed"* by eq.(5), providing, at each step, the distribution of $R(\vec{r},\omega)$ over the relevant wave-front, and therefore the necessary and sufficient condition for the numerical determination of its distribution over the next wave-front.

When, in particular, the space variation length $L$ of the wave amplitude $R(\vec{r},\omega)$ turns out to satisfy the condition $k_0 L >> 1$, the term containing the Wave Potential may be neglected in eqs.(6) and (8). Eq.(6) is well approximated, in this particular case, by the well known "*eikonal equation*" [**19**]

$$(\vec{\nabla}\varphi)^2 \simeq (n\,k_0)^2 \,. \tag{10}$$

In this geometrical optics approximation the rays are no longer mutually coupled by a Wave Potential, and propagate independently from one another under the only influence of the refractive index of the medium. The main consequence of this independence is the absence, in such a limiting case, of typically wave-like phenomena such as diffraction and/or interference, which may only be due to the coupling role of a non-vanishing Wave Potential.

### 3 - The Wave Potential function in Wave Mechanics

Let us pass now to the case of mono-energetic, non-interacting particles of mass $m$ launched with an initial momentum $\vec{p}_0$ into a force field deriving from a stationary potential energy $V(\vec{r})$. The *classical* dynamical behavior of each particle may be described, as is well known [**19**], by the time-independent Hamilton-Jacobi equation

$$(\vec{\nabla} S)^2 = 2\,m\,[E - V(\vec{r})] \,, \tag{11}$$

where $E = p_0^2/2m$ is the total energy of the particle, and the basic property of the function $S(\vec{r},E)$ is that the particle momentum is given by

$$\vec{p} = \vec{\nabla} S(\vec{r},E)\,. \tag{12}$$

In other words, the (time independent) Hamilton-Jacobi surfaces $S(\vec{r},E) = const$ have the basic property of being perpendicular to the momentum of the moving particles, *piloting* them in space along *fixed trajectories*.

One of the most astonishing forward steps in modern physics, giving rise to **Wave Mechanics,** was allowed by **de Broglie's** suggestion [**20**, **21**] that material particles might be associated with suitable "*matter waves*", according to the correspondence

$$\vec{p}/\hbar \equiv \frac{1}{\hbar}\vec{\nabla}S(\vec{r},E) \leftrightarrow \vec{k} \equiv \vec{\nabla}\varphi \qquad (13)$$

assigning to the function $S(\vec{r},E)$ a wave-like nature, but maintaining its *piloting* role and significance. The successive step was accomplished by Schrödinger [**22**, **23**], by assuming that *Classical Mechanics* (represented here by eq.(11)) be the *eikonal approximation* of de Broglie's *Wave Mechanics*, and that his *matter waves* satisfy a *Helmholtz-like equation* of the form (2), whose *eikonal* equation (10) may be written as

$$k^2 \equiv (\vec{\nabla}\varphi)^2 = (n\,k_0)^2 \,. \qquad (14)$$

By performing now, into eq.(14), the replacement

$$\frac{2m}{\hbar^2}[E - V(\vec{r})] \equiv (\vec{\nabla}\frac{S}{\hbar})^2 = p^2/\hbar^2 \rightarrow k^2 \qquad (15)$$

allowed by the relations (11)-(13), we get for the term $(n\,k_0)^2$ an expression which may be inserted into eq.(2), thus obtaining the *time-independent* **Schrödinger** equation

$$\nabla^2 u(\vec{r},E) + \frac{2m}{\hbar^2}[E - V(\vec{r})]\,u(\vec{r},E) = 0 \qquad (16)$$

holding [**24**, **25**] for the *matter waves* associated with mono-energetic particles moving in a stationary potential field $V(\vec{r})$.

Let us remind that the physical existence of *de Broglie's matter waves* was almost immediately confirmed by an experiment performed by Davisson and Germer on electron diffraction by a crystalline nickel target [**26**].

The same mathematical procedure applied in Sect.2 to the Helmholtz eq.(2) may now be applied to the *Helmholtz-like* equation (16), in order to search for a set of exact particle trajectories, analogous to the *ray trajectories* of the previous Section: a search neglected, and even *forbidden*, in the historic development of Wave Mechanics. We put therefore, in analogy with eq.(3) and recalling eq.(13),

$$u(\vec{r},E) = R(\vec{r},E)\,e^{i\,S(\vec{r},E)/\hbar} \,, \qquad (17)$$

and obtain, after separation of real and imaginary parts, the equation systems

$$\begin{cases} \vec{\nabla}\cdot(R^2\,\vec{\nabla}S) = 0 & (18) \\ (\vec{\nabla}S)^2 - 2m(E - V - Q) = 0 & (19) \end{cases} \quad ; \quad \begin{cases} \dfrac{d\vec{r}}{dt} = \dfrac{\partial H}{\partial \vec{p}} \equiv \dfrac{\vec{p}}{m} & (20) \\ \dfrac{d\vec{p}}{dt} = -\dfrac{\partial H}{\partial \vec{r}} \equiv -\vec{\nabla}[V(\vec{r}) + Q(\vec{r},E)] & (21) \end{cases}$$

analogous to (5)-(6) and (7)-(8), respectively, with

$$Q(\vec{r},E) = -\frac{\hbar^2}{2m}\frac{\nabla^2 R(\vec{r},E)}{R(\vec{r},E)} \qquad (22)$$

and

$$H(\vec{r},\vec{p},E) = \frac{p^2}{2m} + V(\vec{r}) + Q(\vec{r},E) = E \,, \qquad (23)$$

where the wave-dynamical Hamiltonian system (20)-(21) - which is seen, by simple inspection, to satisfy the differentiation $\frac{\partial H}{\partial \vec{r}} \cdot d\vec{r} + \frac{\partial H}{\partial \vec{p}} \cdot d\vec{p} = 0$ of eq.(23) - gives us the *exact* particle trajectories and motion laws we were looking for.

The function $Q(\vec{r}, E)$ of eq.(22), which we call once more, for simplicity sake, **"Wave Potential"**, has the same basic structure and role of the **Wave Potential** function $W(\vec{r}, \omega) = -\frac{c}{2k_0} \frac{\nabla^2 R(\vec{r}, \omega)}{R(\vec{r}, \omega)}$ of eq.(9). Although *formally* coincident with Bohm's **statistical** *"Quantum Potential"* $Q_B(\vec{r}, t) = -\frac{\hbar^2}{2m} \frac{\nabla^2 R(\vec{r}, t)}{R(\vec{r}, t)}$ [**7-11**], our **exact** *"Wave Potential"* $Q(\vec{r}, E)$ has a quite different physical role, which shall be stressed in Sect.6.

An important observation, concerning both Bohm's and our own Potentials, is that they have *not so much a "quantum", as a wave-like origin,* entailed into quantum theory by *de Broglie's matter waves*.

The *presence* of the Wave Potential $Q(\vec{r}, E)$ causes, once more, the mutual *"Helmholtz coupling"* (and the wave-like properties) of the trajectories, while its *absence* reduces the system (20)-(21) to the standard set of *classical dynamical* equations, which constitute therefore, as expected [**20-23**], its *geometrical optics* approximation. In *complete analogy* with the electromagnetic case of Section 2,

- because of eq.(18), we have $\vec{\nabla} Q(\vec{r}, E) \cdot \vec{p} = 0$, so that the "force" due to the Wave Potential, acts *perpendicularly*, at each time step, to the particle trajectories, and cannot modify the amplitude of the particle momentum (while modifying, in general, its direction), so that *no energy exchange* is ever involved by the *piloting* action of the *Wave Potential*;

- eq.(18) allows to obtain both $R(\vec{r}, E)$ and $Q(\vec{r}, E)$ along the particle trajectories, thus *providing the "closure" of the quantum-dynamical system* (20)-(21) and making a self-consistent numerical integration possible, without resorting to the simultaneous solution of a time-dependent Schrödinger equation characterizing Bohm's approach.

**4 - Exact particle trajectories versus Uncertainty Principle.**

Many examples of numerical solutions of the Hamiltonian system (20)-(21) in cases of diffraction and/or interference were given in Refs.[**3-6, 18**], in a geometry allowing to limit the computation to the (*x,z*)-plane. The particle trajectories and the corresponding evolution both of *wave intensity* and *Wave Potential* patterns were obtained, *in the absence of external fields*, with initial momentum components

$$p_x(t=0) = 0; \quad p_z(t=0) = p_0 = \hbar_0 k_0 = 2\pi\hbar / \lambda_0, \tag{24}$$

by means of a *symplectic* numerical integration method. We present here, in Fig.1, the *initial* and *final* wave intensity transverse profiles (showing a clear fringe formation) for the diffraction of a *non-Gaussian* particle beam starting from a vertical slit with half-width $w_0$, centered at $x = z = 0$. Fig.2 shows the trajectory pattern corresponding to Fig.1.

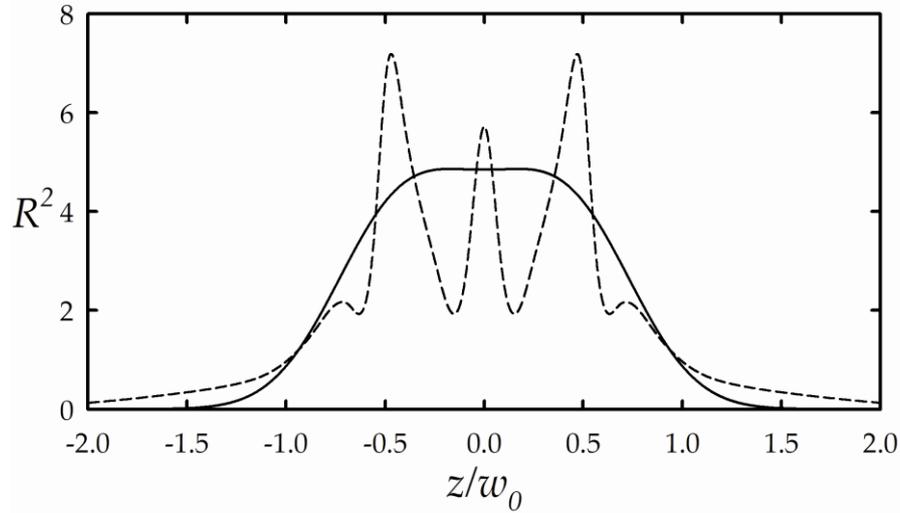

**Fig.1** Initial (continuous) and final (dashed) transverse wave intensity profiles for the diffraction of a *non-Gaussian* beam with $\lambda_0 / w_0 = 2 \times 10^{-4}$ starting from a slit with half-width $w_0$.

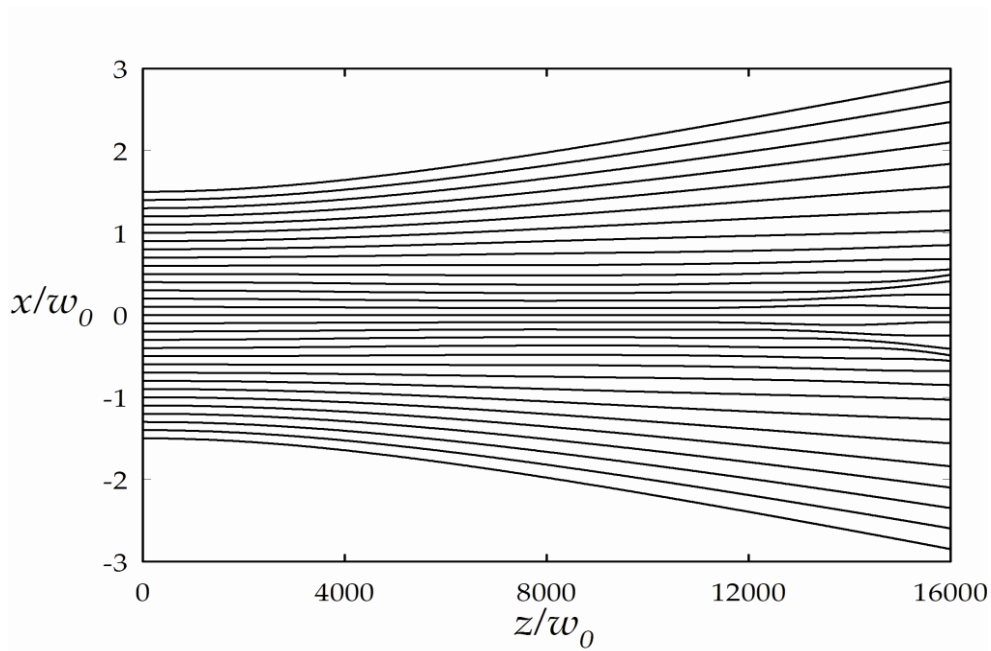

**Fig.2** Wave trajectories corresponding to Fig.1.

Fig.3 shows, in its turn, the *trajectory pattern* obtained in the fringeless diffraction case of a *Gaussian* particle beam of the form $R(x;z=0) \div exp(-x^2/w_0^2)$. The length $w_0$ is the so-called "waist" of the Gaussian beam, and the two heavy lines of Fig.3 represent the so-called *waist-lines* of the beam, given by the well known relation [27]

$$x(z) = \pm \sqrt{w_0^2 + \left(\frac{\lambda_0 z}{\pi w_0}\right)^2} \qquad (25)$$

holding in the quasi-optical *paraxial* approximation and representing the trajectories starting (at $z = 0$) from the positions $x = \pm w_0$.

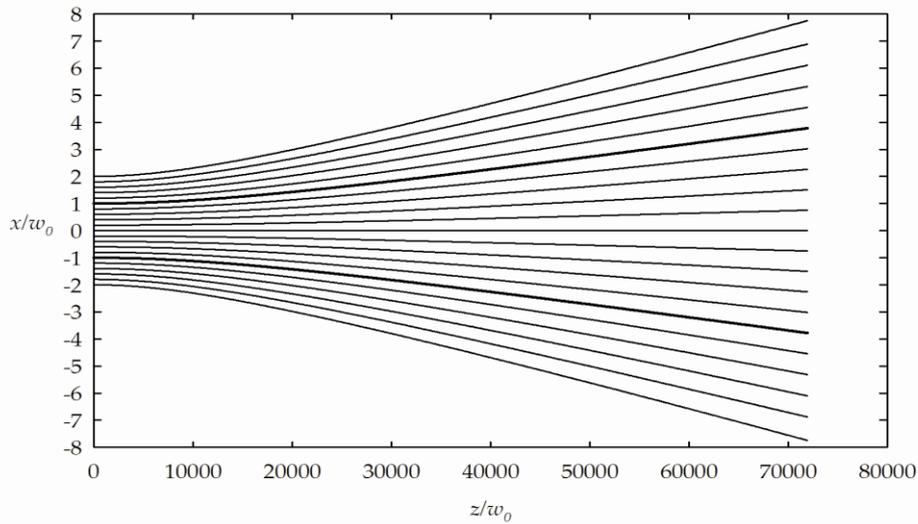

**Fig.3** Wave trajectories and waist lines on the vertical symmetry plane of a *Gaussian* beam with waist $w_0$ and $\lambda_0 / w_0 = 2 \times 10^{-4}$.

The agreement between the *analytical* expression (25) and our *numerical* results provides, of course, an excellent test of our approach and interpretation.

We remind, now, that the passage of a particle through a slit provides a well known procedure for its *space localization*, affected by a *space uncertainty* $\Delta x \approx 2 w_0$. Referring to the Gaussian case, we may observe that, after having crossed the slit, the beam maintains an almost collimated structure, with $p_x \approx 0$, as long as $z << \pi w_0^2 / \lambda_0$, diverging then, for $z >> \pi w_0^2 / \lambda_0$, between the symmetric limiting slopes $\frac{x}{z} \approx \pm \frac{\lambda_0}{\pi w_0}$.

A transverse $p_x(t>0)$ component, ranging between $p_x \approx \pm 2\hbar/w_0$, is therefore progressively developed, under the cumulative action of the Wave Potential, with an uncertainty $\Delta p_x \approx 4\hbar/w_0$, leading to the suggestive *asymptotic* relation

$$\Delta x \, \Delta p_x \approx 8\hbar > h \, , \qquad (26)$$

a relation which is obviously *violated* for $z < \pi w_0^2 / \lambda_0$, i.e. close to the slit location, where $\Delta p_x \approx 0$. The Uncertainty Relation *doesn't appear, therefore, to be a general and intrinsic property of physical reality*, but a local and limited effect - just like diffraction and interference - of the Wave Potential. As in classical mechanics, any possible *space uncertainty* is only due to our lack of knowledge of the starting (point-like) particle position, and the consequent *momentum uncertainty* is an *asymptotic* ($z >> \pi w_0^2 / \lambda_0$) effect of the Wave Potential: an effect which turns out to be negligible in the space region $z < \pi w_0^2 / \lambda_0$. Let us also remind, for completeness sake, that any "disturbance" interpretation of the Uncertainty Principle is nowadays generally recognized [**28**] to have nothing to do with intrinsic reality.

It may be interesting to observe that, in the present approach, such phenomena as diffraction and interference do not concern *particles*, but their (stationary) *trajectories*. Each particle follows, according to the dynamical motion laws (20)-(21), a *fixed* trajectory, ruled by the (stationary) Wave Potential $Q(\vec{r}, E)$; and the overall number of travelling particles is quite indifferent. Speaking, for instance, of self-diffraction (or self-interference) of a single particle is therefore quite inappropriate.

## 5 - Extension to the relativistic case

In order to extend the previous considerations to the relativistic case, we analyze the motion of a particle with rest mass $m_0$, launched, with an energy $E = m(t=0)c^2$, where $m(t=0) = m_0 / \sqrt{1-(\frac{v(t=0)}{c})^2}$, into a force field deriving from a stationary potential energy $V(\vec{r})$. Its behavior may be described by the relativistic *time-independent* Hamilton-Jacobi equation [**24**, **29**]

$$[\vec{\nabla} S(\vec{r},E)]^2 = [\frac{E - V(\vec{r})}{c}]^2 - (m_0 c)^2 \tag{27}$$

which we interpret, once more, as the *geometrical optics approximation* of a *matter wave* satisfying a Helmholtz-like equation of the form (2), whose eikonal equation (10) may be written in the form

$$k^2 \equiv (\vec{\nabla}\varphi)^2 = (n k_0)^2 \quad . \tag{28}$$

By means of eq.(27) we perform therefore the *de Broglie* replacement

$$[\frac{E - V(\vec{r})}{\hbar c}]^2 - (\frac{m_0 c}{\hbar})^2 \equiv [\vec{\nabla} S(\vec{r},E)/\hbar]^2 \equiv \frac{p^2}{\hbar^2} \to k^2 \tag{29}$$

into eq.(28), and insert this expression of the term $(n k_0)^2$ into the Helmholtz eq.(2), which reduces to the *time-independent* **Klein-Gordon** equation [**24**, **29**]

$$\nabla^2 u + [(\frac{E-V}{\hbar c})^2 - (\frac{m_0 c}{\hbar})^2] u = 0 . \tag{30}$$

The use, once more, of eq.(17), followed by the separation of real and imaginary parts, splits now eq.(30) into the equation system

$$\begin{cases} \vec{\nabla} \cdot (R^2 \vec{\nabla} S) = 0 \\ (\vec{\nabla} S)^2 - [\frac{E-V}{c}]^2 + (m_0 c)^2 = \hbar^2 \frac{\nabla^2 R(\vec{r},E)}{R(\vec{r},E)} \end{cases} \tag{31}$$

Making use of the second of eqs.(31), and defining the function

$$H(\vec{r},\vec{p}) \equiv V(\vec{r}) + \sqrt{(pc)^2 + (m_0 c^2)^2 - \hbar^2 c^2 \frac{\nabla^2 R(\vec{r},E)}{R(\vec{r},E)}} = E \tag{32}$$

we may see that its differentiation

$$\frac{\partial H}{\partial \vec{r}} \cdot d\vec{r} + \frac{\partial H}{\partial \vec{p}} \cdot d\vec{p} = 0 \tag{33}$$

is satisfied by the Hamiltonian system ("closed", as usual, by the first of eqs.(31))

$$\begin{cases} \dfrac{d\vec{r}}{dt} = \dfrac{\partial H}{\partial \vec{p}} \equiv \dfrac{c^2 \vec{p}}{E - V(\vec{r})} \\ \dfrac{d\vec{p}}{dt} = -\dfrac{\partial H}{\partial \vec{r}} \equiv -\vec{\nabla} V(\vec{r}) - \dfrac{1}{1 - V(\vec{r})/E} \vec{\nabla} Q(\vec{r}, E) \end{cases} \qquad (34)$$

with

$$Q(\vec{r}, E) = -\dfrac{\hbar^2 c^2}{2E} \dfrac{\nabla^2 R(\vec{r}, E)}{R(\vec{r}, E)}, \qquad (35)$$

providing the relativistic *wave-dynamical* particle trajectories and motion laws, submitted once more to a mutual Helmholtz coupling and reducing to the relativistic *dynamical* description (i.e. to their *eikonal approximation*) in the absence of wave-like effects. Once more, thanks to the first of eqs.(31), the term $\vec{\nabla} Q(r, E)$ acts *perpendicularly* to $\vec{p}$, whose amplitude, therefore, cannot be modified by this wave-like "force".

Somewhat like in the case of a particle with electric charge *e* and relativistic mass *m* moving in *time-independent* electric and magnetic potentials $V(\vec{r})$ and $\vec{A}(\vec{r})$ (a case where $\vec{v} \neq \vec{p}/m$, because of the relation $\vec{p} = m\vec{v} + e\vec{A}(\vec{r})/c$), we have $\vec{v} \equiv \dfrac{d\vec{r}}{dt} \neq \vec{p}/m$ also in the present case, where however $\vec{v} \equiv \dfrac{d\vec{r}}{dt}$ is seen to maintain itself parallel to the momentum $\vec{p}$. It's worthwhile recalling that an expression of the relativistic particle velocity coinciding with the first of eqs.(34) was found by **de Broglie** [**30-33**] in his *double solution theory.*

We conclude the present Section by observing that, in the particular case of *massless* particles (i.e. for $m_0 = 0$) the Klein-Gordon equation (30), by assuming the Planck relation

$$E = \hbar \omega, \qquad (36)$$

takes on the form

$$\nabla^2 u + (\omega n/c)^2 u = 0, \qquad (37)$$

with

$$n(\vec{r}, E) = 1 - V(\vec{r})/E. \qquad (38)$$

Eq.(37) coincides with eq.(2), which may be viewed as the time-independent Klein-Gordon equation of massless point-like particles in a stationary medium. We are therefore brought back to Sect.1 and to the ray trajectories found therein - which present, indeed, interesting analogies with the ones observed in recent *experiments* of single photon interferometry reported in Ref.[**34**].

## 6 - Bohm's Quantum Potential

Coming now to a comparison with Bohm's approach, let us previously recall that, starting from eqs.(1) and (16), one gets

$$\nabla^2 \psi - \dfrac{2m}{\hbar^2} V(\vec{r}) \psi = -\dfrac{2m}{\hbar^2} E \psi \equiv -\dfrac{2mi}{\hbar} \dfrac{E}{\hbar \omega} \dfrac{\partial \psi}{\partial t}, \qquad (39)$$

an equation which, by assuming the Planck relation (36), reduces to the usual form of the *time-dependent* Schrödinger equation for a stationary potential field $V(\vec{r})$:

$$\nabla^2 \psi - \frac{2m}{\hbar^2} V(\vec{r}) \psi = -\frac{2mi}{\hbar} \frac{\partial \psi}{\partial t}, \qquad (40)$$

where $E$ and $\omega$ are not explicitly involved, and no wave dispersion is therefore, in principle, described. As is well known [24, 25], indeed, eq.(40), representing a rare example of intrinsically complex equation in physics, is not even a wave equation: its wave-like implications are only due to its connection with the *time-independent* eq.(16).

While, starting from the *time-independent* Schrödinger equation eq.(16), the *time-dependent* eq.(40) is a mathematical truism, its "stronger" version:

$$\nabla^2 \psi - \frac{2m}{\hbar^2} V(\vec{r},t) \psi = -\frac{2mi}{\hbar} \frac{\partial \psi}{\partial t}, \qquad (41)$$

containing a time-dependent potential, $V(\vec{r},t)$, may only be accepted (and is in fact generally accepted) as an **assumption**. Bohm's approach [7-11] performs, as is well known, a replacement of the form

$$\psi(\vec{r},t) = R(\vec{r},t)\, e^{\,i\,S(\vec{r},t)/\hbar} \qquad (42)$$

into eq.(41) itself, splitting it, after separation of real and imaginary parts, into the equation system

$$\begin{cases} \dfrac{\partial P}{\partial t} + \vec{\nabla}\cdot(P\dfrac{\vec{\nabla} S}{m}) = 0 \\[6pt] \dfrac{\partial S}{\partial t} + \dfrac{(\vec{\nabla} S)^2}{2m} + V(\vec{r},t) - \dfrac{\hbar^2}{2m}\dfrac{\nabla^2 R}{R} = 0 \end{cases} \qquad (43)$$

where, in agreement with the standard Copenhagen interpretation, the function $P \equiv R^2$ is assumed to represent, in Bohm's own terms, the *probability density for particles belonging to a statistical ensemble*. The **first** of eqs.(43) is viewed as a *fluid-like* continuity equation for such a probability density, and the **second** of eqs.(43) is viewed, in its turn, as analogous to a Hamilton-Jacobi *dynamical* equation, containing however - strangely enough - a *statistical* "Quantum Potential"

$$Q_B(\vec{r},t) = -\frac{\hbar^2}{2m}\frac{\nabla^2 R(\vec{r},t)}{R(\vec{r},t)}, \qquad (44)$$

to be compared with our *exact* Wave Potential $Q(\vec{r},E) = -\dfrac{\hbar^2}{2m}\dfrac{\nabla^2 R(\vec{r},E)}{R(\vec{r},E)}$ of eq.(22).

A further Ansatz (the so-called "**guidance assumption**") is also performed, in the form $\vec{\nabla} S \equiv \vec{p}$, suggested by this *dynamical* analogy: an analogy from which Bohm infers that "*precisely definable and continuously varying values of position and momentum*" [7] may be associated, *in principle*, with each particle. Since however, "*the most convenient way of obtaining R and S is to solve for the* (time-dependent) *Schrödinger wave function*", we are led, *de facto*, to an **unavoidably statistical** *description* of the particle motion.

The situation is even more evident if we limit our attention to a *stationary* external potential $V(\vec{r})$. In this case the *time-independent* Schrödinger equation (16) admits in general, as is well known [**24**, **25**], a (discrete or continuous, according to the boundary conditions) set of energy eigen-values and orthonormal eigen-modes which (referring for simplicity to the discrete case) we shall call, respectively, $E_n$ and $u_n(\vec{r})$.

If we make use of eqs.(1) and (36), and define both the eigen-frequencies $\omega_n \equiv E_n/\hbar$ and the eigen-functions

$$\psi_n(\vec{r},t) = u_n(\vec{r})\, e^{-i\omega_n t} \equiv u_n(\vec{r})\, e^{-iE_n t/\hbar} \quad , \tag{45}$$

together with a (duly normalized) linear superposition (with constant coefficients $c_n$) of the form

$$\psi(\vec{r},t) = \sum_n c_n \psi_n, \tag{46}$$

we may easily verify that such a superposition provides a general solution of the *time-dependent* Schrödinger equation (40).

The wave-packet (46) is, in any case, a weighted, time-evolving sum performed over the full set of the solutions $u_n$ of the **time-independent** Schrödinger equation (16), where the relative weights $c_n$ are determined by the available physical information. Any feature due to mono-energetic properties (such as transverse trajectory coupling) is dimmed, therefore, by this average character, which hinders the possibility of distinguishing their individual peculiarities: the description, for instance, of *diffraction* and *interference* features - requiring, in order to be observed, a good amount of monochromaticity - imposes, for a wave packet, a very careful choice of the set of parameters $c_n$, strictly centering it around a particular eigenfunction $\psi_n$.

## 7 - Discussion

The solution (46) owes its fame, as is well known, to Born's *ontological interpretation* [**35**] as the expression of **a physical state** where energy is not determined: an interpretation which, even though *"no generally accepted derivation has been given to date"* [**36**], has become one of the standard principles of Quantum Mechanics and a new philosophical conception of physical reality, in spite of the host of "quantum paradoxes" it raises and, to say the least, of Ockham's razor. One could argue, indeed, that the unusual, intrinsically complex nature of the **time-dependent** Schrödinger equation (40) is basically connected with the fact that *it doesn't describe,* after all, the evolution of a physical state, but of an average of states.

We would like to stress, once and for all, that the **time-independent** Schrödinger equation (16), as well as the **time-independent** Klein-Gordon equation (30), are a direct result of de Broglie's hypothesis (13), and that these equations are **the origin**, and **not the consequence**, of their **time-dependent** partners, inextricably accompanied by Born's probabilistic interpretation. Since, in spite of this fact, standard opinions conceive *the successes of the **time-independent** equations as particular cases of the **time-dependent** ones*, it's high time to reverse the current point of view. To be sure,

- our *exact, stationary, energy-dependent function Q($\vec{r}$,E) stemming from the **time-independent** Schrödinger equation,* turns out to exert a consistently dynamical action on classical-looking *point-like particles*, while

- Bohm's *statistical, time-evolving, energy-independent* function $Q_B(\vec{r},t)$ acts, on the contrary, on "particles" represented as *wave-packets,* i.e. as statistical averages evolving along probability flux-lines *according to the **time-dependent** Schrödinger equation*.

Needless to say, our point-like particle trajectories conflict - both in the relativistic and in the non-relativistic case - with the idea (induced by the Copenhagen interpretation) that the very concept of " trajectory" is physically meaningless. Although this conflict is already present, *in principle*, in Bohm's *hidden-variables* theory, where the idea of "trajectories" is admitted, Bohm refrains, *in practice*, from a full responsibility assumption by representing particles by means of *wave-packets*, i.e. of statistical ensembles which are declared to be "*a practical necessity"*, and by repeatedly stressing the full equivalence between his own results and the "orthodox" ones.

Our approach performs, instead, a *further, crucial step*, pulling those "*hidden variables*" out of their hiding place and referring to point-like particles without any probabilistic contraption. While Bohm's approach doesn't appear to differ so much from the standard Copenhagen paradigm, with which it associates a set of *fluid-like* probability flux-lines [**37**] representing an average over a set of *conjectured* exact trajectories, we explicitly *determine* these exact trajectories, providing the *fine-grained* theoretical structure which underlies Bohm's *coarse-grained* description.

## 8 - Conclusion

We mention here a reflection due to E.T. Jaynes [**38**]:

"*Our present quantum mechanical formalism is not purely epistemological; it is a peculiar mixture describing in part realities of Nature, in part incomplete human information about Nature - all scrambled up by Heisenberg and Bohr into an omelette that nobody has seen how to unscramble. Yet we think that this unscrambling is a pre-requisite for any further advance in basic physical theory*".

Clearly enough, Bohm's probability flux lines, giving a "visual" representation of the standard solution of the time-dependent Schrödinger equation, correspond to an average (a "*scrambling*") taken over the *exact* trajectories found in the present work. We may conclude that our *non-probabilistic* approach represents the *wave-mechanical* "missing link" between the description provided by classical (both relativistic and non-relativistic) particle dynamics and the Copenhagen (and Bohm) *statistical* description, thus satisfying Jaynes' "*unscrambling pre-requisite for a further advance in basic physical theory*", and suggesting a *novel, non probabilistic* interpretation of Wave Mechanics.